# Eliminating Shadow Artifacts in OCT via Generative Inpainting Networks to Quantify Vascular Changes of the Choroid

Huihong Zhang[1,5], Jianlong Yang[1,*], Kang Zhou[1,4], Fei Li[2], Yan Hu[3], Shenghua Gao[4], Xiulan Zhang[2], and Jiang Liu[1,3]

[1] Ningbo Institute of Industrial Technology, Chinese Academy of Sciences, China

[2] State Key Laboratory of Ophthalmology, Zhongshan Ophthalmic Center, Sun Yat-sen University, China

[3] Southern University of Science and Technology, China

[4] ShanghaiTech University, China

5 University of Chinese Academy of Sciences, China

* Corresponding Author: Jianlong Yang (yangjianlong@nimte.ac.cn )

## Abstract:

**Purpose:** To avoid the inaccurate quantification of choroidal vasculature caused by the projected retinal vessel shadows in OCT, we proposed a deep-learning-based image processing framework that could eliminate the shadow artifact from the choroid. We verify the feasibility of the proposed framework using an observational study that detects the choroidal vascular changes related to the intraocular pressure (IOP) elevation.

**Methods:** A swept-source system (Atlantis, DRI OCT-1; Topcon, Japan) was employed for data acquisition. We took 6×6×2 mm$^3$ volumetric scans of 68 eyes of 34 healthy volunteers. The shadow locations and the en face choroid were inputted into a two-stage generative inpainting network which connects the edges of the vessels ahead of refilling the shadow-contaminated areas. Then we converted the shadow-free choroid into binary vessel map and reflectance-based flow map for quantifying vessel density and a relative flow index. To simulate the state of high IOP, after taking the baseline scans in a normal sitting position, each of the volunteers was asked to take scans in the upside-down position.

**Results:** Qualitative and quantitative results show the proposed image processing framework is capable of completely eliminating the vessel shadows while keeping the morphology of the underneath choroid.

**Conclusions:** The deep-learning-based shadow removal frame improved the quantification accuracy of the choroid vasculature thus benefitted the extraction of the vascular biomarkers.

**Translational Relevance:** It is promising to further apply this method in the studies of pathological changes of the choroid.

# Introduction

The choroid, lying between the retina and the sclera, is the vascular layer which provides oxygen and nourishment to the outer retina[1]. Traditional imaging modalities like fundus camera and ophthalmoscope acquire 2D overlapping information of the retina and the choroid, from which choroidal vascular information cannot be extracted separately so that the vascular changes of the choroid could not be precisely retrieved and evaluated. Optical Coherence Tomography (OCT) is a non-invasive 3D imaging modality that could separate the information of the underlying choroid from the retina, thus has been becoming a powerful tool to understand the role of the choroid in various ocular diseases[2]. It has been shown that the thickness of the choroid layer extracted from OCT, especially the subfoveal choroid thickness, is directly related to the incidence and severity of predominate ocular diseases, such as pathologic myopia[3], Diabetic Retinopathy (DR)[4], Age-related Macular Degeneration (AMD)[5], and glaucoma[6].

On the other hand, the choroidal vasculature, which could also be acquired from volumetric OCT data, has relatively limited applications in the study and diagnosis of ocular diseases[7-8]. The contamination induced by the vessel shadows from the superficial layers of the retina is the primary reason that limits the extraction of quantitative vascular biomarkers from the underneath layers[9]. As shown on the left side of Fig. 1, the anisotropy of the red blood cells inside the vessels cause strong forward scattering of the probe light, thus bring shadow-like dark tails to the underneath layers extending to the choroid and the sclera (inside the dashed red boxes). The center part of Fig. 1 is a layer-segmented OCT volume, which could further be used to generate the *en face* images of each layer as shown in the right of Fig. 1. Inside the green box is the Ganglion Cell Layer (GCL) which possesses the retinal vessels with high light reflectance. While the depth-projected vessel positions turn dark on the vessel-absent Retinal Pigment

Epithelium (RPE) layer (inside the red box) and the choroid layer (inside the orange box). It is evident that the shadows bring difficulties to the extraction of the choroidal vasculature.

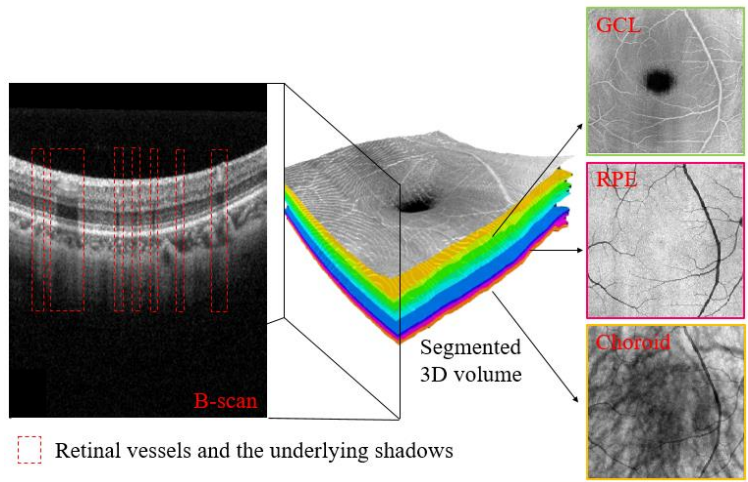

Figure 1. Demonstration of the shadows of the retinal vessels and their influences on the underlying layers especially the choroidal vasculature.

In 2011, Girard et al. developed an attenuation compensation (AC) algorithm to remove the OCT vessel shadows and enhance the contrast of optic nerve head[10]. This algorithm was then employed in the calculation of the attenuation coefficients of retinal tissue[11], enhancing the visibility of lamina cribrosa[12], and improving the contrast of the choroid vasculature and the visibility of the sclera-choroid interface[13-14]. However, the shadow removal capacity of this algorithm was relatively limited to those from small vessels and varies among the OCT images from different commercial OCT systems as shown in Fig. 2. We can see the vessel shadows are mostly minimized in the images from Zeiss CIRRUS and Heidelberg SPECTRALIS systems but largely uninfluenced in the images from Topcon Atlantis system. The red arrows shows the remaining shadows after applying the AC algorithm.

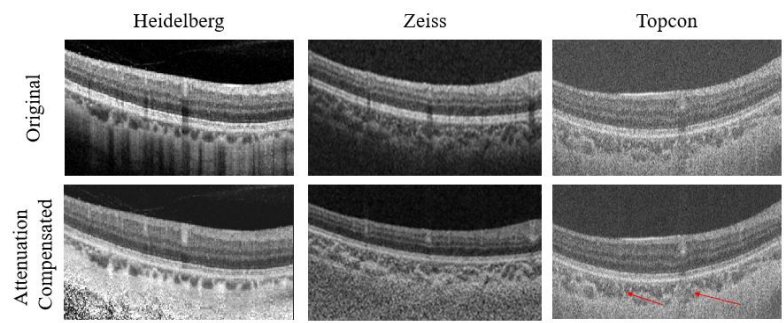

Figure 2. Performance of the AC algorithm on the OCT images from different commercial systems.

To universally eliminate the contamination of the retinal vessel shadows to the choroid, we proposed a deep-learning-based image processing framework. The OCT volume data was first enhanced with the AC algorithm. Because of its avascular nature, the RPE layer was segmented to locate the vessel shadows (As shown in Fig. 1). The choroid was segmented and mean-value projected into the *en face* plane. The shadow locations and the *en face* choroid were inputted into a two-stage generative inpainting network which connects the edges of the vessels ahead of refilling the shadow-contaminated areas. Then we converted the shadow-free choroid into binary vessel map and reflectance-based flow map for quantifying vessel density and a relative flow index. We further applied the proposed method in the application of quantitatively detecting the changes of the choroidal vasculature in the intra-ocular pressure (IOP) elevation. The results show it is prospective in studying the pathological changes of the choroid.

## Methods

### Study Population

Written informed consent was obtained from all participants involved in the study before the recruitment. The participants were 34 healthy volunteers with the ages ranging from 18 to 30 years old with no previous history of IOP exceeding 21 mmHg. The study was approved by the Ethical Review Committee of the Zhongshan Ophthalmic Center and was conducted in accordance with the Declaration of Helsinki for research involving human subjects.

### Data Acquisition

We employed a swept source OCT system (Atlantis, DRI OCT-1; Topcon, Japan) to collect data from both of their eyes. We used the $6\times6$ mm$^2$ Field Of View (FOV) volumetric scan protocol centered at the fovea, which was composed of 256 B-scans and 512 A-lines in each B-scan. Each A-line has 992 data points uniformly distributed in a depth range of ~ 2 mm. To simulate the state of high IOP, each of the volunteers was asked to take scans in the upside-down position after taking the baseline scans in a normal sitting position. The average IOP was increased to 34.48±5.35 mmHg because of the upside-down, compared with the average IOP of 15.84 ±1.99 mmHg at the normal position. The IOP was measured using a portable tonometer (iCare, Tiolat Oy; Helsinki, Finland).

## Deep learning based image processing framework

The proposed framework is composed of three stages to process the volumetric OCT data, as shown in Fig. 3.

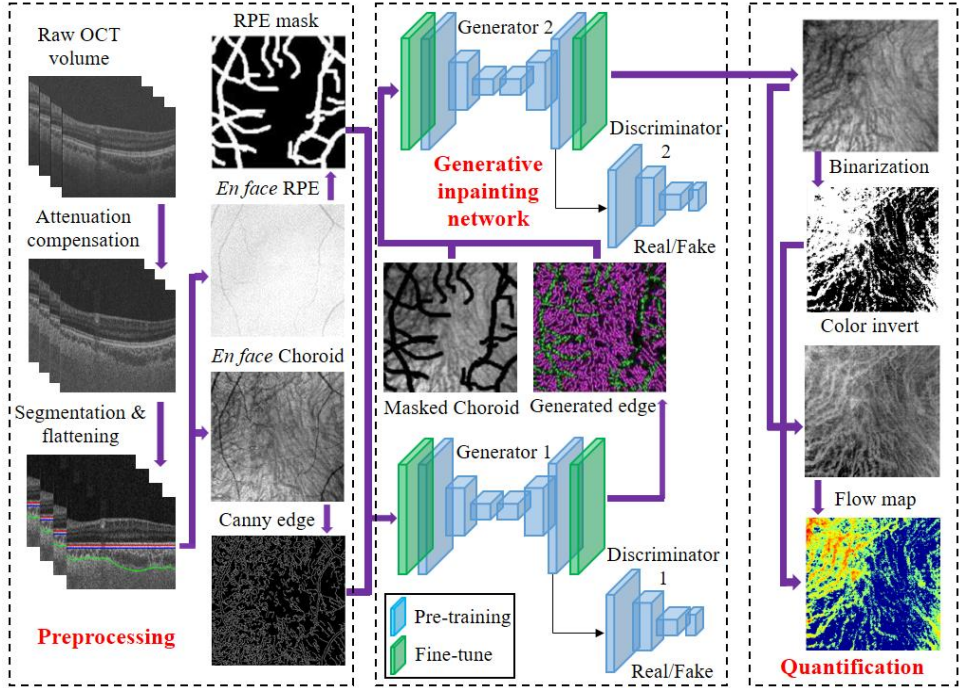

Figure 3. Graphic illustration of the proposed framework, which is composed of three stages. Stage 1: Preprocessing the raw OCT volume to acquire the RPE mask and Canny edge. Stage 2: Eliminating the retinal vessels shadow via Generative Inpainting Network. Stage 3: Quantifying the vascular changes of the choroid.

*Preprocessing*

The boundary between the choroid and the sclera in the OCT data was firstly enhanced via the AC algorithm[10], which could improve choroidal vessel contrast and minimizing the retinal shadows[12-14]. It can be expressed as:

$$I_{AC}(x,y) = \frac{I(x,y)}{2\sum_{k=x}^{M} I(k,y)}, \qquad (1)$$

Where $I_{AC}$ is the AC-enhanced image intensity and $I$ is the original image intensity. $(x, y)$ are the pixel coordinates of the B-scan images ($x \in [1, M]$, $y \in [1, N-1]$). $M$ and $N$ are the row and column numbers. This boundary and the boundary between the RPE and Bruch's membrane were automatically segmented with a graph search algorithm[15]. Then the OCT volume was flatten based on the RPE. We obtained the *en face* RPE and choroid by calculating axial mean projection value among these boundaries. Finally, the vessel shadows on the avascular RPE image was used to extract the mask and the choroid image was used to extract the edge using the Canny method[16].

*Generative Inpainting Network*

Recently, Generative Adversarial Networks (GAN)[17] has demonstrated its superiority in handling image inpainting tasks compared with traditional algorithms[18-20]. We propose to use a two-stage GAN[21] for the shadow removal task. The first stage takes masked canny edge $\mathbf{E}_m$ as input, while the second stage takes masked choroid $\boldsymbol{C}_m$ as input. Let $\mathbf{E}_{gt}$ and $\mathbf{C}_{gt}$ be ground-truth of the Canny edge and choroid respectively, while $\mathbf{E}_{pred}$ and $\mathbf{C}_{gt}$ denote output prediction of the first and second stages, respectively. The first stage takes $\mathbf{E}_m$ as the input and eliminates mask with the edge generator $\boldsymbol{G}_e$, $\mathbf{E}_{pred} = \boldsymbol{G}_e(\mathbf{E}_m)$, It is trained with an edge adversarial loss and Mean Absolute Error (MAE) loss as follows:

$$\min_{\boldsymbol{G}_e} \max_{\boldsymbol{D}_e} \mathcal{L}_{G_e} = \min_{\boldsymbol{G}_e}\left(\lambda_{adv,1}\max_{\boldsymbol{D}_e} \mathcal{L}_{adv,1} + \lambda_{mae,1}\|\boldsymbol{G}_e(\mathbf{E}_m) - \mathbf{E}_{gt}\|_1\right), (2)$$

$$\mathcal{L}_{adv,1} = \mathrm{E}[\log \mathbf{D}_e(\mathbf{E}_{gt})] + \mathrm{E}\left[\log\left(1 - \mathbf{D}_e(\mathbf{G}_e(\mathbf{E}_m))\right)\right], \quad (3)$$

where $\lambda_{adv,1}$ and $\lambda_{mae,1}$ are regularization parameters.

The second stage takes $\boldsymbol{C}_m$ with $\mathbf{E}_{pred}$ as input and eliminates mask with generator $\mathbf{G}_c$, $\mathbf{C}_{pred} = \mathbf{G}_c(\boldsymbol{C}_m, \mathbf{E}_{pred})$. It is trained with an adversarial loss and MAE loss as follows:

$$\min_{\mathbf{G}_c} \max_{\mathbf{D}_c} \mathcal{L}_{\mathbf{G}_c} = \min_{\mathbf{G}_c}\left(\lambda_{adv,2}\max_{\mathbf{D}_c} \mathcal{L}_{adv,2} + \lambda_{mae,2}\|\mathbf{G}_c(\boldsymbol{C}_m, \mathbf{E}_{pred}) - \boldsymbol{C}_{gt}\|_1\right),$$

(4)

$$\mathcal{L}_{adv,2} = E[\log \boldsymbol{D}_c(\boldsymbol{C}_{gt})] + E\left[\log\left(1 - \boldsymbol{D}_c\left(\mathbf{G}_c(\boldsymbol{C}_m, \mathbf{E}_{pred})\right)\right)\right], \quad (5)$$

Where $\lambda_{adv,2}$ and $\lambda_{mae,2}$ are regularization parameters. For our experiments, we empirically choose $\lambda_{adv,1} = 1$, $\lambda_{mae,1}= 10$. $\lambda_{adv,2} = 0.1$ and λadv;2 = 0:1 and $\lambda_{mae,2} = 1$.

*Quantification*

After eliminating retinal vessels shadow, we segmented choroidal vasculature by thresholding and binarizing. The generated vessel map was used for calculating the vessel density [22]. We inverted the intensity value in the *en face* choroid. Then it was combined with the vessel map to get the flow map, which could be further used to calculate the flow index [22].

$$vessel\ density = \frac{\int_A V dA}{\int_A dA}, \tag{7}$$

$$flow\ index = \frac{\int_A I \cdot V dA}{\int_A dA}, \tag{6}$$

Where A is the area used in the calculation. I is the intensity of the *en face* choroid image. If the pixel belongs to a vessel, $V = 1$, otherwise $V = 0$.

*Image Quality Assessment*

Mean Squared Error (MSE), Structure Similarity Index (SSIM), and Peak Signal to Noise Ratio (PSNR) were employed as quantitative metrics. They are calculated as:

$$MSE = \frac{\Sigma \|I_{pred} - I_{gt}\|^2}{\Sigma \|I_{gt}\|^2}, \tag{8}$$

where $I_{pred}$ and $I_{gt}$ denote the predicted image and the ground truth image respectively.

$$PSNR = 10\ log_{10} \frac{MAX^2}{\frac{1}{MN}\Sigma \|I_{pred} - I_{gt}\|^2}, \tag{9}$$

where MAX is the maximum possible pixel value of the image. M and N are the scale (rows and columns) of the input image.

$$SSIM(x, y) = \frac{(2\mu_x \mu_y + c_1)(2\sigma_{xy} + c_2)}{(\mu_x^2 + \mu_y^2 + c_1)(\sigma_x^2 + \sigma_y^2 + c_2)}, \tag{10}$$

where $(x, y)$ is the pixel coordinate, $x \in (0, M - 1), y \in (0, N - 1)$. $\mu_x$ and $\mu_y$ are the average of $x$ and $y$. $\sigma_x^2$ and $\sigma_y^2$ are the variance of $x$ and $y$. $c_1 = (0.01T)^2$ and $c_2 = (0.03T)^2$ with T the maximum value allowed for the data.

## Results

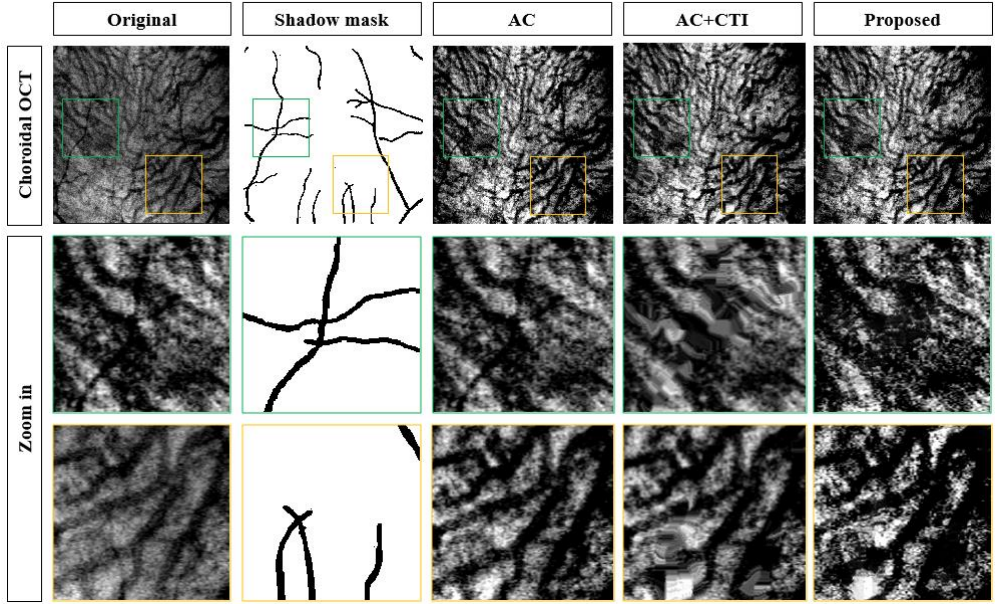

Figure 4. Comparison of the proposed method with the prior arts. AC: attenuation compensation. CTI: coherence transport inpainting.

The results using the proposed framework were compared with the AC algorithm. We also included another inpainting-based shadow removal approach using traditional coherence transport inpainting (CTI)[23] for comparison. As shown in Fig. 4, the original choroidal vasculature is contaminated by the retinal vessel shadows at the locations shown in the shadow mask. Inside the green and/or yellow boxes are two selected zoom-in regions. The AC could enhance the contrast of the choroidal vessels and minimize faint shadows as shown in the yellow box. However, it cannot get rid of the dark shadows as shown in the green box. With the assistance of the inpainting-based approach, both the faint and dark shadows could be thoroughly eliminated. However, the CTI introduces unnatural artifacts, as shown in the zoom in boxes.

Table 1. Quantitative comparison of different inpainting algorithms.

| Methods / Metrics | Masked | EBI | CTI | Proposed | |
|---|---|---|---|---|---|
| | | | | Pre-trained | Fine-tuned |
| SSIM ↑ | 0.778±0.085 | 0.901±0.043 | 0.931±0.015 | 0.918±0.039 | **0.946±0.013** |
| PSNR ↑ | 11.851±2.605 | 25.988±3.117 | 29.352±3.712 | 29.990±3.253 | **30.615±3.824** |
| MSE (×10$^3$) ↓ | 3.464±1.895 | 0.165±0.092 | 0.109±0.082 | 0.105±0.079 | **0.091±0.071** |

Due to the absence of the ground-truth choroidal vasculature, we created the artificial retinal vessel mask with the vessel widths slightly wider than the real shadows (the retinal vessels in the dataset are less than 8 pixels in a $512 \times 512$ pixels *en face* image, we used the vessel widths of 10 to 15 pixels). The artificial mask combining with the repaired choroid images were used as the input to evaluate the inpainting algorithms. Besides the CTI and the proposed GIN, we also included the classical Exemplar-Based Inpainting (EBI) algorithm[24] in the evaluation. Structure Similarity Index (SSIM), Peak Signal to Noise Ratio (PSNR), and Mean Squared Error (MSE) were employed as quantitative metrics. We used the masked images as the baseline. The results were summarized in Table 1. We tested these metrics using their default functions in MATLAB. The SSIM measurements in Table 1 suggest the good feasibility of the proposed inpainting-based method in this task. All of the testing methods have the SSIMs of >0.9, which implies a realistic restoration of the choroidal vasculature. We got high MSE values because the images were not normalized to [0 1], which would not influence the conclusion of the comparison. In accordance with the experimental results above, the finetuned GIN has the best performances in the quantitative of all the metrics. The CTI outperformed the pre-trained GIN, which suggests the discrepancy of the *en face* choroidal images compared with the natural scene images used in the pre-trained model.

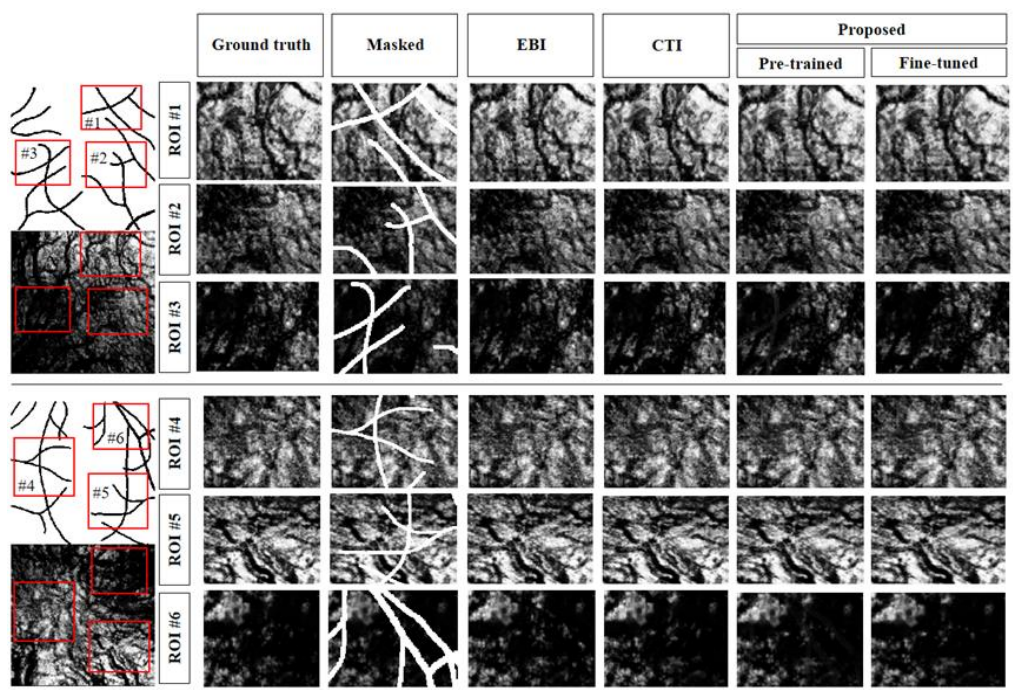

Figure 5. Quantitative comparison of different inpainting algorithms. EBI: exemplarbased inpainting. CTI: coherence transport inpainting. ROI: region of interest

Figure 5 shows the results using different algorithms. Six Region Of Interests (ROIs) from two *en face* choroid images are used for demonstration. Based on visual inspection, the traditional algorithms tend to induce more inpainting artifacts than the GIN. But all of them could repair the missing areas with good visual similarity with the ground truth.

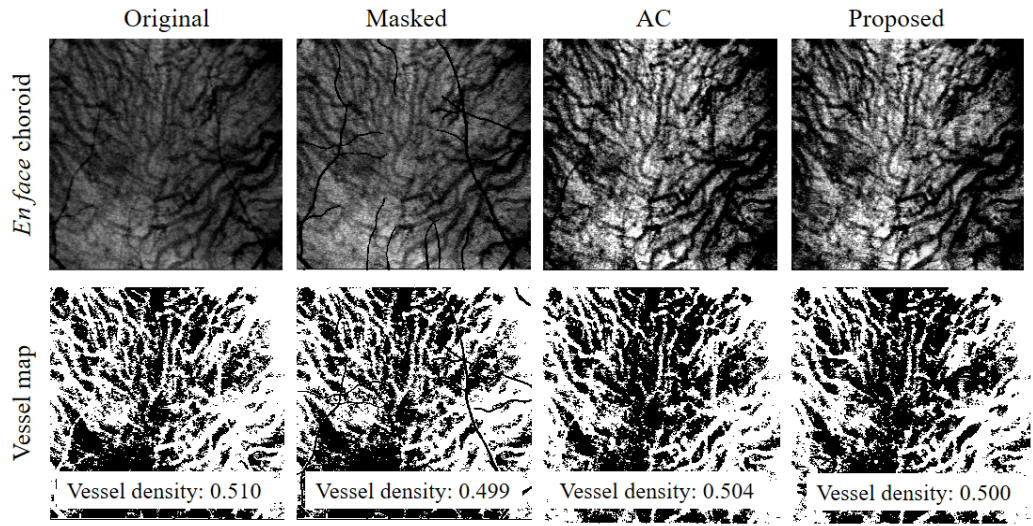

Figure 6. Application of the proposed method in the quantitative measurement of choroidal vessel density.

It is necessary to further exam the validity of the proposed shadow removal method in quantifying the choroidal vasculature. We employed the calculation of the vessel density as an example. The vessel maps were generated using the *en face* choroid images from the original data, the shadow-masked image, the AC-processed image, and the one using the proposed method, respectively as shown in Fig. 6. We calculated the vessel density values from each of them. We employed the shadow-masked image as the ground-truth of this comparison. The contamination of the vessel shadows was completely avoided because the shadow areas were not included in the vessel density calculation. We can see the original images overestimate the vessel density because the shadows were also recognized as the choroidal vasculature. The AC algorithm did minimize the shadows from the small retinal vessels, but the large vessel shadows still exist. The vessel density value is smaller than the original image but still larger than that of the ground truth. The proposed method, on the other hand, got a much closed vessel density value to the ground truth. It can be seen the shadows has been completely removed in the vessel map.

After removing the retinal shadows, the choroidal vasculature can be clearly visualized and used in the quantitative analysis. Figure 7 demonstrates the calculated vessel and flow maps before and during the upside down of two subjects in the datasets. Obvious decrease in vessels and blood flow could be observed in the images of the high IOP state, which further induces the reduction in the mean values. In the datasets, the choroidal vessel density changes from $0.491 \pm 0.020$ at the normal state to $0.463 \pm 0.019$ during the high IOP state. Their corresponding flow indexes are $0.336 \pm 0.025$ and $0.300 \pm 0.019$, respectively.

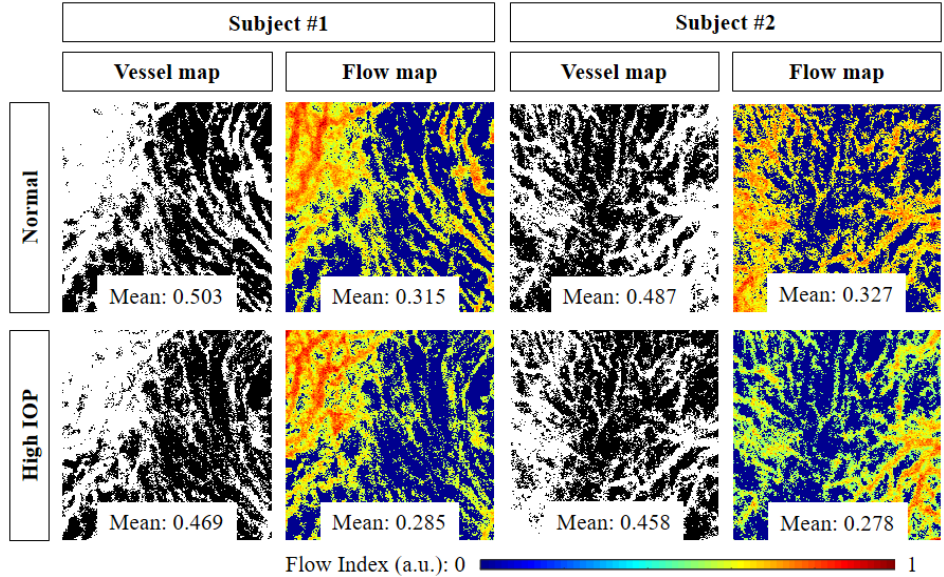

Figure 7. Application of the proposed method in the quantitative measurement of choroidal vasculature changes related to high IOP.

## Discussion and Conclusions

We have developed a deep-learning-based image processing framework, which could efficiently eliminate the contamination of the retinal vessel shadows to the choroid in OCT. The shadow artifacts would lead to inaccurate estimation of the choroidal vasculature biomarkers, such as the overestimation of the vessel density as shown above. The previous AC algorithm could only remove the small vessel shadows but have limited effects on the large vessel shadows. We have employed the avascular RPE layer to locate both the small the large vessel shadows, thus the elimination is complete. Compared with the traditional object removal methods, the adopted deep-learning-based GIN recover the shadow-contaminated areas without generating artifacts. We have further validated this method in detecting the vessel and blood flow deduction caused by the high IOP. It shows excellent perspectives in pathological researches and clinical applications, such as glaucoma and pediatric myopia.

However, the proposed method still has two major limitations. (1) It may not suitable for those diseases with an abnormal RPE layer, such as drusen and CNV. Under the circumstances, the shadow locations used for the inpainting task are hard to extract. One alternative solution may be to directly locate the retinal vessels or shadows in OCT with advanced image processing algorithms. Its feasibility has been demonstrated[25] but needs the extra information provided by fundus photos, which brings difficulty and complexity. (2) It may not be suitable for extending the proposed method in projection-resolved OCT angiography (OCTA)[26]. It is natural to think using this method to minimize the projection artifacts on intermediate capillary plexus (ICP), deep capillary plexus (DCP), and chorocapillaries (CC) in OCTA, but we are cautious about this extended application. Because the morphological scales of the choroidal vasculature are larger than the retinal vessels and shadows, the recovery of the contaminated areas have shown to be realistic as shown in Fig. 7 and Table 1. However, the ICP, DCP, and CC have morphological scales of less than the superficial vessels shadows, the inpainting process may bring unrealistic vessel information.

At the mention of OCTA, a recent paper[27] explains why we not use it in this study. It compared the imaging of the choroid using OCT and OCTA and found that for the CC, the vessel density as measured by OCTA was significantly greater than that measured by OCT. However, for the choroidal, the advantages of OCTA disappear. In the inner choroid, OCTA did not appear to capture the choroidal vasculature as fully as OCT, which led to consistent underestimation of the vessel density. In the outer choroid, vessel lumens appeared larger in OCTA, which led to consistent overestimation of the vessel density as compared with OCT.

In conclusion, we have proposed a method to eliminate the retinal shadow artifacts in volumetric OCT data for the quantitative analysis of the choroidal vasculature. The shadow removal was converted into an inpainting task and handled with a two-stage generative adversarial network. With the specially-designed data set and experiments, we have evaluated the feasibility and performance of the proposed framework. The results show that this work is promising in paving the road to apply quantitative vascular biomarkers of the choroid in the study and diagnosis of various ocular diseases.

## Acknowledgments

Ningbo 3315 Innovation team grant; Cixi Institute of Biomedical Engineering, Chinese Academy of Sciences (Y60001RA01, Y80002RA01); Zhejiang Provincial Natural Science Foundation (LQ19H180001); Ningbo Public Welfare Science and Technology Project (2018C50049).

The authors declare no conflicts of interest.